\documentclass[aps,prl,nofootinbib,floatfix,twocolumn,showpacs]{revtex4-1}

\usepackage{pstricks}
\usepackage{graphicx}
\usepackage{bm}
\usepackage{amsmath,amssymb,amsfonts}
\begin{document}

\title{Twisting wire scanner}

\author{V. Gharibyan}
\email[]{vahagn.gharibyan@desy.de}
\author{A. Delfs}
\author{I. Krouptchenkov}
\author{D. Noelle}
\author{H. Tiessen} 
\author{M. Werner} 
\author{K. Wittenburg} 

\affiliation{Deutsches Elektronen-Synchrotron DESY - D-22603 Hamburg}

\begin{abstract}
A new type of 'two-in-one' wire scanner is proposed. Recent advances in linear motors' technology make it possible to combine translational and rotational movements. This will allow to scan the beam in two perpendicular directions using a single driving motor and a special fork attached to it. Vertical or horizontal mounting will help to escape problems associated with
the 45 deg scanners. Test results of the translational part with linear motors is presented.
\end{abstract}

\pacs{ 41.85.Ew, 29.27.Fh, 07.07.Tw }
\maketitle

\section{INTRODUCTION}

Wire scanners serve as an essential part of accelerator diagnostic systems
and are used mostly for beam transverse  profile
measurements (for a review see~\cite{Wittenburg:2006zz}). Depending on scanning wire
trajectory the profilers could be classified as rotational~\cite{Fischer:1988ft}
or linear~\cite{Loos:2010zzc}. When its necessary to measure
vertical and horizontal beam profiles at the same longitudinal position
one has to use two independent scanners. Alternatively two profiles could be sampled
by using a single driver mounted at 45deg with two wires stretched horizontally and
vertically over a fork attached to this linear driver.
However, wire vibration in the scanning direction is a known problem for the 45deg
scanners~\cite{Frisch:2008zz, Iida:1998ua}.
\begin{figure}[htb]
   \centering
   \includegraphics*[width=85mm]{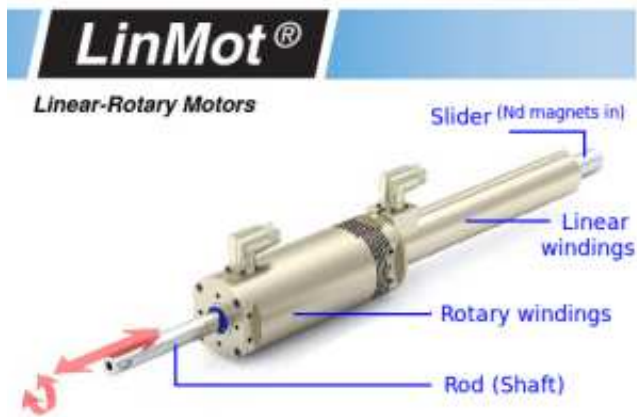}
\caption{Linear-rotary motor from  LinMot company.  }
   \label{lrmot}
\end{figure}
   Different types of driver motors have been employed
in order to move and control scanning wires which are normally mounted on cards or
forks connected to the motors. Stepper or servo rotating motors are among the most
popular drivers and linear motors are at developing stage.
Here we explore commercially available translational-rotational motor units
to propose a wire scanner solution which will perform beam scans in mutually
perpendicular directions using a single linear-rotary motor and a simple
wire hosting construction attached to it. The construction is a key-like
wire holder which makes twisting (helical) motion during a 2-D scan.
Next will follow a more detailed description of the translational part with linear
motors. In conclusion we will estimate technical feasibility of the proposed twisting scanner.

\section{LINEAR-ROTARY MOTORS}

A linear-rotary motor produced by company LinMot~\cite{linmot} is shown in Fig.~\ref{lrmot}.
The motor consists of a linear and a rotary part merged together.
Translational and rotational motions are decoupled and organized independently.
However, linear and rotary motion synchronization is foreseen by motor controller logic.
The motors are provided in different configurations with variable sizes and strengths
reaching up to $1~kN$ linear force and $7.5~Nm$ rotating torque.
\begin{table}[hbt]
   \centering
   \setlength\tabcolsep{2pt}
   \caption{LinMot PR01-52x60-R/37x120F-HP-100 Linear-rotary Motor Parameters}
   \begin{tabular}{ll}
       \toprule
        \textbf{Parameter} & \textbf{Value} \\ \hline
\multicolumn{2}{c}{Linear Motion} \\ 
Extended Stroke ES mm (in) & 100 (3.94)\\
Standard Stroke SS mm (in) &100 (3.94)\\
Peak Force E12x0 - UC N (lbf)& 255 (57.3)\\
Cont. Force N (lbf) &51 (11.5)\\
Cont. Force Fan cooling N (lbf)& 92 (20.7)\\
Force Constant N/A (lbf/A) &17 (3.8)\\
Max. Current @ 72VDC A& 15\\
Max. Velocity  m/s (in/s) &3.9 (154)\\
Position Repeatability mm (in) & $\pm0.05$ ($\pm0.0020$)\\
Linearity \% &  $\pm0.10$\\
\multicolumn{2}{c}{Rotary Motion} \\ 
Peak Torque Nm (lbfin)& 2 (17.7)\\
Constant Torque (Halt) Nm (lbfin) &0.5 (4.4)\\
Max. Number of revolutions Rpm& 1500\\
Torque\,Constant\,Nm/Arms\,(lbfin/Arms)& 0.46 (4.07)\\
Max. Current @ 72VDC Arms& 6.2\\
Repeatability $\deg$ &$\pm0.05$\\
   \end{tabular}
   \label{linmot}
\end{table}
Motor controllers
use advanced and flexible software/firmware which should help to perform slow or fast scans
with minimal programming efforts. An operational voltage of 72VDC and maximal current
of 15A complies to general Electro-Magnetic Interference (EMI) requirements in accelerator
environments. Described features make the linear-rotary motor as an attractive tool for
driving the proposed twisting  wire scanner.
A closer look to specifications of a linear-rotary motor  LinMot
\hbox{PR01-52x60-R/37x120F-HP-100}
is presented in Table~\ref{linmot} as an example.
Listed values for the Repeatability are quoted for built-in, internal position and angle
sensors. One can improve these parameters considerably by using external, finer sensors
which is foreseen by controller software. In following we demonstrate that for linear motors.

\section{KEY-BIT SCANNER}

\begin{figure}[htb]
   \centering
   \includegraphics*[width=85mm]{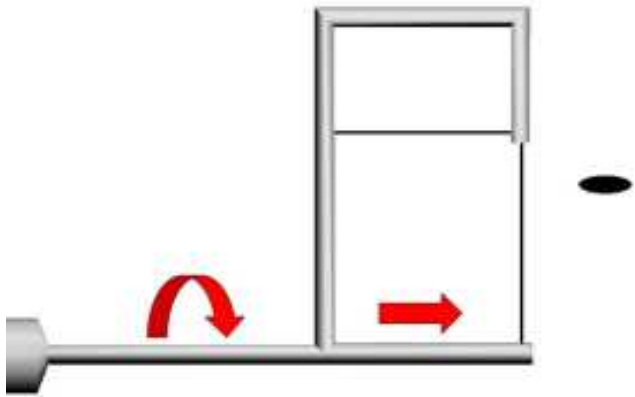}
\caption{A key-bit holder scheme with horizontal and vertical scanning wires.
A small ellipse on the right depicts a beam running normal to the page.}
   \label{fork}
\end{figure}

In order to apply 2-D helical motion of a linear-rotary motor for scanning a
beam, one needs to invent a suitable construction with stretched wires and a
holding frame which stays out of (does not cross the) beam during the scan.
For that we propose a key-bit like assembly which fulfills above requirements.
The construction is schematically presented in Fig.~\ref{fork}.
As it's indicated by arrows, for this  arrangement translational motion will first
scan the beam in horizontal direction and next, when the beam will be inside the key-bit,
a proper rotation will perform vertical scan.  It is necessary to limit the rotation angle
in order to escape crossing of the wire holder with the beam. For that there is sufficient
space between the holder frame and the beam, remaining after the rotational scan is over.
Denoting vertical key-bit and wire size by $L$ and $l_w$ respectively, the beam to holder
distance could be expressed as
$$ L\left( \arccos{\frac{x}{L}} - \arccos{\frac{x}{l_w}}  \right) $$
where $x$ is distance between the beam and rotational axis. Applying this formula for
fast ($>1~m/sec$) rotational scans with some realistic accelerator parameters we
obtain sufficient space to accelerate the wire while for deceleration the space is
limited and one needs to use mechanical dumps to stop the scanner.

\section{3-D KEY-BIT HOLDER FOR 2-D FAST SCANS}

An improved, slightly more complicated design, for the fast scans could be achieved
\begin{figure}[htb]
   \centering
   \includegraphics*[width=75mm]{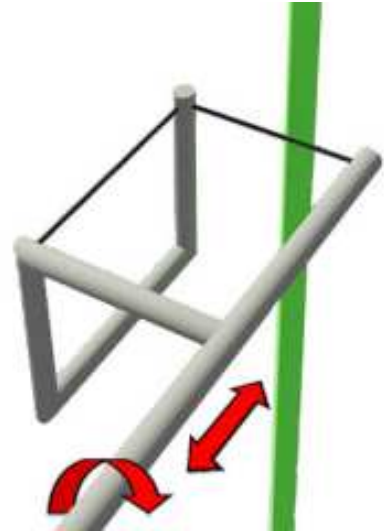}
\caption{Scanning scheme of a three dimensional key-bit wire holder. Vertical green
line depicts the beam to be scanned.}
   \label{fork3d}
\end{figure}
by tilting the second quadrant of the key-bit wire holder out of the construction's plane
by some angle. A 90 deg tilted key-bit holder is sketched in Fig.~\ref{fork3d}.
Scanning sequence is exactly the same as for the flat key-bit scanner with an advantage
of more room after the second scan is over.
 This should give sufficient time to decelerate
and stop the frame by the motor alone, without  mechanical dumpers.  In addition the tilted
 key-bit's moment of inertia   is considerably smaller than in flat case.
This will allow easier and improved handling of rotations with more ergonomic acceleration
and deceleration of the key-bit structure.

\section{LINEAR MOTOR PERFORMANCE AS A SCANNER DRIVER}

\begin{figure}[htb]
   \centering
   \includegraphics*[width=70mm]{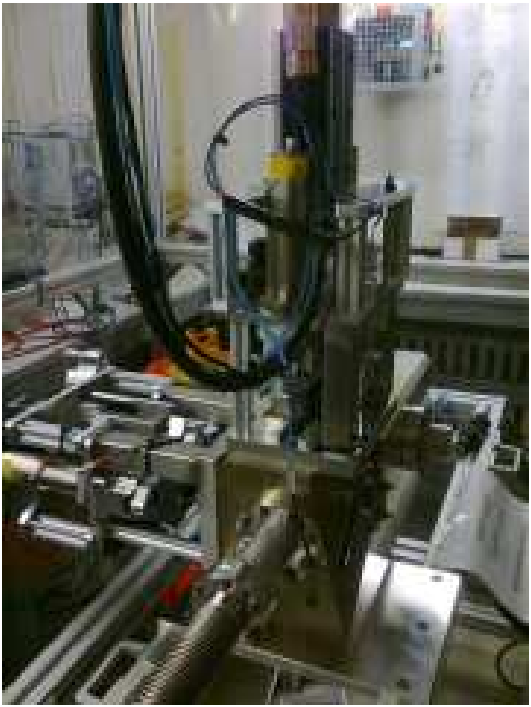}
\caption{XFEL wire scanners' test setup }
   \label{test1}
\end{figure}

We are developing wire scanners with linear motor drivers for European XFEL accelerator.
\begin{figure}[htb]
   \centering
   \includegraphics*[width=95mm]{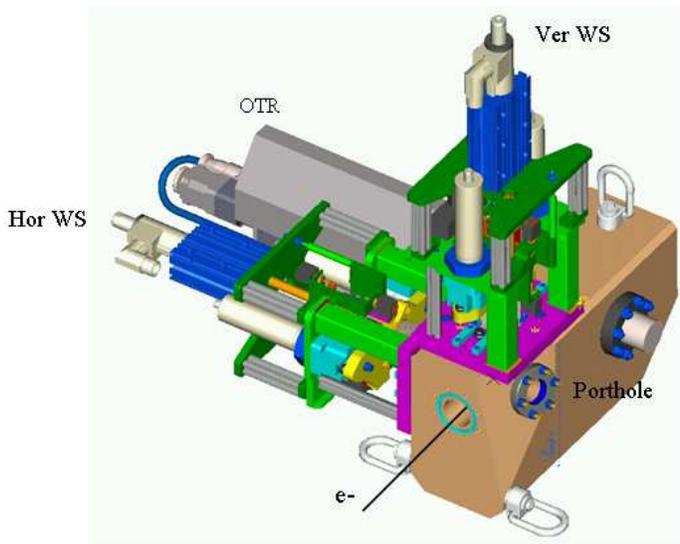}
\caption{Designed horizontal and vertical wire scanners mounted on vacuum chamber.  }
   \label{xfel3d}
\end{figure}
Here we present some of the results obtained during recent laboratory experiment with
test scanners. Planned test setup is displayed in Fig.~\ref{xfel3d} while experimental
realization is presented by Fig.~\ref{test1}. For horizontal and vertical scans two
identical and independent profilers are mounted to a special vacuum chamber dedicated
to beam transverse diagnostics.

Position feedback for the linear servo-motor is provided by an external Heidenhain optical
system which is accurate to $1\mu m$.
With the setup we have tested triggered fast
scans and mechanical as well linear motors performance during/after tens of thousand scanning
strokes.
Important specifications of the XFEL wire scanners are shown in Table~\ref{xfel-tab}.
\begin{table}[hbt]
   \centering
   \caption{European XFEL Wire Scanner Specifications}
   \smallskip
   \begin{tabular}{ll}
       \toprule
        \textbf{Parameter} & \textbf{Value} \\ \hline
Stroke & 53mm \\
Measurement  duration &5 sec / 4 scanners \\
Scanning modes& Fast (1m/s), Slow\\
Motor to  beam sync  &   $< 1 \mu sec$ (RMS)\\
Position accuracy in a cycle &2 $\mu m$  (RMS)\\
Width accuracy per cycle & 2 \%    (RMS)  \\
Wire positioning error   &   1 $\mu m$\\
Number of wires per fork & 3 + 2 ( 3x$90^o$,  $\pm 60^o$)\\
Wire material & Tungsten \\
Fork gap & 15mm\\
Wire-wire distance  & 5mm ( $90^o$ ) \\
   \end{tabular}
   \label{xfel-tab}
\end{table}
Tests have marked most of the listed specifications as achieved.
During the test mechanical design and construction precision has been justified
while linear motors have demonstrated reliable performance.
\begin{figure}[htb]
   \centering
 \includegraphics*[width=\columnwidth]{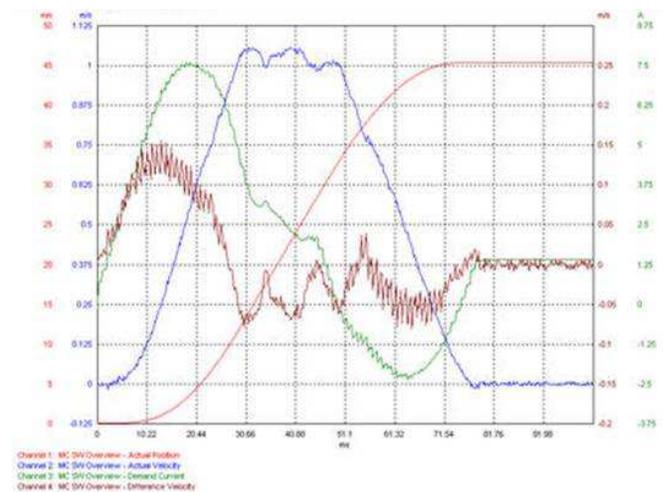}
\caption{Linear motor parameters recorded during a stroke: position(mm, red),
velocity(m/s, blue), current (A, green), demand and actual velocity difference (m/s, brown) }
   \label{motpar}
\end{figure}
To verify motor's dynamic behavior we have recorded essential parameters during nominal
strokes. An example is shown on Fig.~\ref{motpar} where together with position and velocity
also the motor's current and velocity deviation are displayed for a fast ($1m/s$) scanning stroke.
An important issue for the XFEL wire profilers and fast scans in general is mechanical jitter
magnitude for triggered scans. We have investigated this by recording time intervals between
the trigger and fine position system reference mark traversing time.
\begin{figure}[htb]
   \centering
 \includegraphics*[width=\columnwidth]{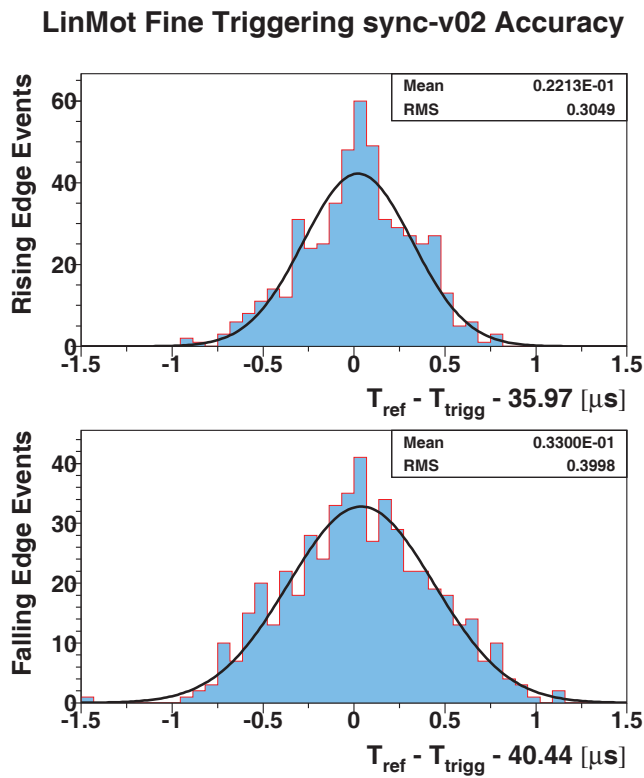}
\caption{Motor triggering mechanical jitter distribution for
 the forward (upper plot) and backward (lower plot) strokes. Superimposed are
shown fitting gaussian functions with displayed RMS (one sigma) values. }
   \label{trigg}
\end{figure}
Measurement results for
many forward and backward strokes are summarized in Fig.~\ref{trigg}.
Distributions show time jitter below $1 \mu sec$ which, in our case of $1 m/sec$
velocity, is equivalent to a sub-micrometer mechanical jitter.
This could also be quoted as a repeatability of the tested linear motor with
fine position feedback and triggering systems.

\section{DISCUSSION}

In the last section we have demonstrated an outstanding performance of contemporary
linear motors as wire scanner drivers.
We have proposed to use linear-rotary motor with attached key-bit wire card as
2-D twisting wire scanner. Estimated planes of possible oscillations of the key-bit
wires differ from critical  planes in 45~deg forks which  should cure
associated vibrational problems reported at LCLS and other centers. This will become
possible mainly due to different alignment of the scanning wires relative to
driver unit. In addition the vibrations are normally dumped along the motion direction.

An apparent difficulty for twisting wire scanner development is the linear-rotary motion
transfer into the vacuum chamber where the key-bit card should operate.
For that one should combine linear bellows with either wobble~\cite{Bosser:1984us} or
torsional~\cite{tbellows} bellows.

\newpage

\acknowledgements{
We are thankful to  S.~Vilcins-Czvitkovits, J.~Kruse, A.~Ziegler and  M.~Drewitsch for
participation in preparation and running of the XFEL wire scanners' first prototyping
test.}

\end{document}